\RequirePackage{fix-cm}
\documentclass[twocolumn,epjc3]{svjour3}
\smartqed  
\RequirePackage{graphicx}
\usepackage[usenames, dvipsnames]{color}
\usepackage{lmodern}
\usepackage{amsmath}
\newcommand{\be}{\begin{equation}}
\newcommand{\ee}{\end{equation}}
\newcommand{\bn}{\begin{eqnarray}}
\newcommand{\en}{\end{eqnarray}}
\newcommand{\bes}{\begin{subequations}}
\newcommand{\ees}{\end{subequations}}

\usepackage{epstopdf}
\usepackage{widetext}
\usepackage{amssymb}
\usepackage{hyperref}
\journalname{Eur. Phys. J. C}
\begin{document}

\title{Wormholes in $R^2$-gravity within the $f(R,T)$ formalism}


\author{P.K. Sahoo$^{1,a}$
\and P.H.R.S. Moraes$^{2,b}$
\and Parbati Sahoo  $^{1,c}$
}

\thankstext{1}{e-mail: pksahoo@hyderabad.bits-pilani.ac.in}
\thankstext{2}{e-mail: moraes.phrs@gmail.com}
\thankstext{3}{e-mail: sahooparbati1990@gmail.com}

\institute{Department of Mathematics, Birla Institute of Technology and Science-Pilani, Hyderabad Campus, Hyderabad-500078, India
\and
ITA - Instituto Tecnol\'ogico de Aeron\'autica - Departamento de F\'isica, 12228-900, S\~ao Jos\'e dos Campos, S\~ao Paulo, Brazil
}

\date{Received: date / Accepted: date}

\maketitle

\begin{abstract}

We propose, as a novelty in the literature, the modelling of wormholes within the particular case of the $f(R,T)$ gravity, namely $f(R,T)=R+\alpha R^{2}+\lambda T$, with $R$ and $T$ being the Ricci scalar and trace of the energy-momentum tensor, respectively, while $\alpha$ and $\lambda$ are constants. Although such a functional form application can be found in the literature, those concern to compact astrophysical objects, such that no wormhole analysis has been done so far. The quadratic geometric and linear material corrections of this theory make the matter content of the wormhole to remarkably be able to obey the energy conditions.

\keywords{$f(R,T)$ gravity \and wormholes \and energy conditions}
\end{abstract}

\section{Introduction}
\label{sec:int}

Recent cosmological observations indicate that the universe is undergoing an accelerated expansion \cite{Perlmutter/1999,Riess/1998}. The cause of it still remains an open and tantalizing question for the research community. Theoretically, several works are propose to resolve this problem. Modifications in standard gravitational theory play a significant role once they are an alternative to explain the accelerated expansion. 

One of the first and simplest modifications to Einstein-Hilbert action is named $f(R)$ gravity, with $f(R)$ being a function of the Ricci scalar. In the literature, this theory has a lot of contributions that account for the dark energy problem. A. A. Starobinsky has proposed a viable dark energy model within the framework of $f(R)$ gravity, by assuming $f(R)=R+\alpha R^{2}$, with constant $\alpha$ \cite{starobinsky/1980}. It was shown in such a reference that the inception of the $R^2$ term removes some weak singularities. This was further developed in References \cite{Thongkool/2009,Appleby/2010} (check also the review \cite{Nojiri/2011}). Moreover, the inflationary era and dark matter issues are also addressed in this theory \cite{Myrzakulov/2015}-\cite{Bamba/2015}.

In 2011, the modified $f(R)$ gravity was generalized from the inception of a function of the trace of the energy-momentum tensor in the gravitational action, yielding the $f(R,T)$ gravity \cite{harko/2011}, which is going to be worked in the present paper. It has been studied in several astrophysical and cosmological aspects, by assuming different choices for the arbitrary function of $R$ and $T$, as it can be seen in \cite{Houndjo/2012}-\cite{Shabani/2017}, for instance. 

In the present paper, we will consider $f(R,T)=f_1(R)+f_2(T)$, with $f_1(R)=R+\alpha R^{2}$ and $f_2(T)=\lambda T$, where $\alpha$ and $\lambda$ are arbitrary constants. The choice of $f_1(R)$ is considered as given by the Starobinsky model \cite{starobinsky/1980} (check also \cite{Starobinsky/2007}). The Starobinsky model has gained importance in recent years for its role in the analysis of matter density perturbations, inflation and many other applications \cite{Fu/2010}-\cite{Kaneda/2016}. Here, we are going to analyse wormhole (WH) solutions from the Starobinsky model within the $f(R,T)$ gravity. 

WHs are two-way tunnels, attaching different regions of space-time. The minimal surface area of that attachment is known as the ``throat'' of the WH. M.S. Morris and K.S. Thorne, which firstly suggested these objects \cite{morris/1988}, showed that matter inside them has negative energy, that violates the null energy condition (NEC). Anyhow, in the last few years, there has been a growing interest in developing some exact WH models that account for the minimization of the violation or even the obedience of NEC, which in fact can be attained in modified gravity theories (check, for instance, \cite{Garcia/2011,jawad/2016}). 

WH solutions can be seen in the literature in the framework of $f(R,T)$ gravity in different aspects. In Ref.\cite{Azizi/2013}, WHs were firstly analysed within the linear frame of $f(R,T)$ gravity, i.e., $f(R,T)=R+2\lambda T$, with the matter Lagrangian $\mathcal{L}_m=-\rho$ and $\rho$ the matter-energy density of the WH. Static numerical solutions have been obtained for different WH matter contents and stable solutions were attained in \cite{Zubair/2016,Yousaf/2017}. Moraes et al. have investigated some theoretical predictions for static WHs obtained from $f(R,T)$ gravity where the matter Lagrangian is governed by the total pressure of the WH \cite{mcl/2017}. In \cite{ms/2017b}, a static WH model has been constructed from various shape functions . In \cite{Yousaf/2017a}, the authors have studied spherical WH models with different fluid configurations. Motivated by the above references, we are going to check here if it is possible to obtain stable WH solutions in $R^{2}$-gravity within the $f(R,T)$ formalism. 

It is worth to remark that the functional form to be used here, named $f(R,T)=R+\alpha R^2+\lambda T$, has already been applied to the analysis of compact astrophysical objects in References \cite{noureen/2015}-\cite{zubair/2015}, but no WH analysis has been made so far for such a functional form. In these references, it was shown that such a model is well motivated since it is consistent with stable stellar configurations because the second-order derivative with respect to $R$ remains positive for the assumed choice of the parameters and this prior choice agrees with the energy conditions outcomes of the present article. Moreover, Starobinsky has shown that his model (free from the $T$-dependence) predicts an overproduction of scalarons in the very early universe \cite{Starobinsky/2007}. This issue was also addressed, for instance, in \cite{koshelev/2016b,gorbunov/2015}. On the other hand, it has been shown that the presence of the trace of the energy-momentum tensor of a scalar field in a gravity formalism can well address the inflationary era \cite{ms/2016}. So, the $T$-dependence inserted in the Starobinsky formalism  indeed raises as a promising model of gravity to be deeply and widely investigated. 

\section{The $f(R,T)$ gravity model}\label{sec:frt}

The action in the $f(R,T)$ theory of gravity is given by \cite{harko/2011}
\begin{equation}\label{1}
S=\frac{1}{16\pi}\int d^{4}x\sqrt{-g}f(R,T)+\int d^{4}x\sqrt{-g}\mathcal{L}_m,
\end{equation}
where $f(R,T)$ is an arbitrary function of the Ricci scalar $R=R^i_j$ and trace of the energy-momentum tensor $T=T^i_j$, $g$ is the metric determinant and $\mathcal{L}_m$ is the matter Lagrangian density. Here, we have assumed the speed of light $c$ and gravitational constant $G$ to be $1$.

By varying the action $S$ with respect to the metric $g_{ij}$, the following $f(R,T)$ field equations are obtained \cite{harko/2011}

\begin{multline}\label{2}
R_{ij}f_R(R,T)-\frac{1}{2}f(R,T) g_{ij}+(g_{ij}\square-\nabla_i\nabla_j)f_R(R,T)=\\
8 \pi T_{ij}-f_T(R,T)\theta_{ij}-f_T(R,T)T_{ij}.
\end{multline}
Here, we are using the notations $f_R(R,T)\equiv\partial f(R,T)/\partial R$ and $f_T(R,T)\equiv\partial f(R,T)/\partial T$, $T_{ij}$ is the energy-momentum tensor and  $\theta_{ij}$ is defined as

\begin{equation}\label{3}
\theta_{ij}=g^{\alpha \beta}\frac{\delta T_{\alpha \beta}}{\delta g^{ij}}=-2T_{ij}+g_{ij}\mathcal{L}_m-2g^{\alpha \beta}\frac{\partial^2 \mathcal{L}_m}{\partial g^{ij} \partial g^{\alpha \beta}}.
\end{equation}

In this paper, we take the energy-momentum tensor for the relativistic source of anisotropic fluid to be in the form

\begin{equation}\label{4}
T_{ij}=(\rho+p_t)u_i u_j-p_t g_{ij}+(p_r-p_t)X_i X_j,
\end{equation}
where $\rho, p_r,$ and $p_t$ are the energy density, radial pressure and tangential pressure, respectively. Moreover $u_i$ and $X_i$ are the four-velocity vector and radial unit four-vector, respectively, which satisfy the relations $u^{i} u_i=1$ and $X^i X_i=-1$. 

By considering the matter Lagrangian as $\mathcal{L}_m=-\mathcal{P}$, Equation (\ref{3}) can be rewritten as
\begin{equation}\label{5}
\theta_{ij}=-2T_{ij}-\mathcal{P} g_{ij},
\end{equation}
where $\mathcal{P}=\frac{p_r+2p_t}{3}$ is the total pressure, such that the trace of the energy-momentum tensor reads $T=\rho-3\mathcal{P}$. Hence, the $f(R,T)$ gravity field equations (\ref{2}) with (\ref{5}) take the form

\begin{equation}\label{6}
R_{ij}-\frac{1}{2}Rg_{ij}= T_{ij}^{eff},
\end{equation}
where $T_{ij}^{eff}$ is the effective energy-momentum tensor of the theory, defined as

\begin{multline}\label{7}
 T_{ij}^{eff}=\frac{1}{f_R(R,T)}\left\{[8\pi+f_T(R,T)]T_{ij}+\mathcal{P}g_{ij}f_R(R,T)\right\}\\
 + \frac{1}{f_R(R,T)}\left\{\frac{1}{2}[f(R,T)-Rf_R(R,T)]g_{ij}\right\}\\
-\frac{1}{f_R(R,T)} \left[(g_{ij}\square-\nabla_i\nabla_j)f_R(R,T)\right].
\end{multline}

\section{Wormhole solutions in $R^2$-gravity within the $f(R,T)$ formalism}\label{sec:ws}

The static spherically symmetric WH metric in Schwarzschild coordinates $(t,r,\theta, \phi)$ reads \cite{morris/1988,visser/1995}

\begin{equation}\label{8}
ds^2=e^{2\phi(r)}dt^2-\frac{dr^2}{1-\frac{b(r)}{r}}-r^2(d\theta^2+sin^2\theta d\phi^2),
\end{equation}
where $\phi(r)$ and $b(r)$ are the redshift and shape functions, respectively. The radial coordinate $r$ increases from a minimum radius value $r_0$ to $\infty$, i.e, $r_0\leq r <\infty$, where $r_0$ is known as the WH throat radius. A flaring out condition of the throat is considered as an important condition to have a typical WH solution, such that $\frac{b-b'r}{b^2}>0$  \cite{morris/1988,visser/1995} and at the throat, $r=r_0=b(r_0)$ and $b'(r_0)<1$, with primes denoting radial derivatives. Another condition needed to be respected by the shape function is $1-\frac{b(r)}{r}>0$ \cite{morris/1988,visser/1995}. To ensure the absence of horizons and singularities, the redshift function $\phi(r)$ needs to be finite and non-null throughout the space-time. In this work we will consider the redshift function as a constant, for simplicity, and also to achieve the de Sitter and anti-de Sitter asymptotic behaviour of the metric \cite{cataldo/2011}.

The effective field equations (\ref{6}) for the metric (\ref{8}) and our specific functional form for the $f(R,T)$ function, i.e, $f(R,T)=R+\alpha R^2+\lambda T$, read

\begin{equation}
\frac{b'}{r^2}=\frac{\left[\left(8\pi+\frac{3\lambda}{2}\right)\rho-\frac{\lambda(p_r+2p_t)}{6}-\frac{2\alpha b'^2}{r^4}\right]}{2\alpha R+1},\label{9}
\end{equation}
\begin{equation}
\frac{b}{r^3}=\frac{\left[-\left(8\pi+\frac{7\lambda}{6}\right)p_r+\frac{\lambda}{2}\left(\rho-\frac{2p_t}{3}\right)-\frac{2\alpha b'^2}{r^4}\right]}{2\alpha R+1},\label{10}
\end{equation}
\begin{equation}
\frac{b'r-b}{2r^3}=\frac{\left[-\left(8\pi+\frac{4\lambda}{3}\right)p_t+\frac{\lambda}{2} \left(\rho-\frac{p_r}{3}\right)-\frac{2\alpha b'^2}{r^4}\right]}{2\alpha R+1}.\label{11}
\end{equation}

From the above equations, the explicit form for the matter quantities, i.e., $\rho$, $p_r$ and $p_t$, can be written as follows:

\begin{equation}\label{12}
\rho=\frac{ b' \left[\lambda  \left(2 r^2-5 \alpha   b'\right)+12 \pi  \left(r^2-2 \alpha   b'\right)\right]}{3 (\lambda +4 \pi ) (\lambda +8 \pi ) r^4},
\end{equation}

\begin{widetext}
\begin{equation}\label{13}
p_r=-\frac{-12 \alpha  b \lambda  b'-48 \pi  \alpha  b b'-\lambda  r^3 b'+7 \alpha  \lambda  r b'^2+24 \pi  \alpha  r b'^2+3 b \lambda  r^2+12 \pi  b r^2}{3 (\lambda +4 \pi ) (\lambda +8 \pi ) r^5},
\end{equation}

\begin{equation}\label{14}
p_t=-\frac{12 \alpha  b \lambda  b'+48 \pi  \alpha  b b'+\lambda  r^3 b'+12 \pi  r^3 b'+2 \alpha  \lambda  r b'^2-3 b \lambda  r^2-12 \pi  b r^2}{6 (\lambda +4 \pi ) (\lambda +8 \pi ) r^5}.
\end{equation}
\end{widetext}

The exact solutions for the matter content of the WH can be obtained by considering a specific form for the shape function, as \cite{Heydarzade/2015}

\begin{equation}\label{16}
b(r)=r_0+ar_0\left[\left(\frac{r}{r_0}\right)^{\beta}-1\right] ,\,\,\,\,\ 0<\beta<1,
\end{equation}
where $a$ is an arbitrary constant. The flaring out condition $\frac{b-b'r}{b^2}>0$ implies that $1-a+a\left(\frac{r}{r_0}\right)^{\beta}(1-\beta)>0$.

Similarly, at the throat, we have $a \beta<1$ for $b'(r_0)<1$. Also, $\frac{b(r)}{r}=(1-a)\frac{r_0}{r}+a\left(\frac{r_0}{r}\right)^{1-\beta}\rightarrow 0$ when $r\rightarrow \infty$ for $\beta<1$. 

\begin{figure}[ht!]
\centering
  \includegraphics[width=75mm]{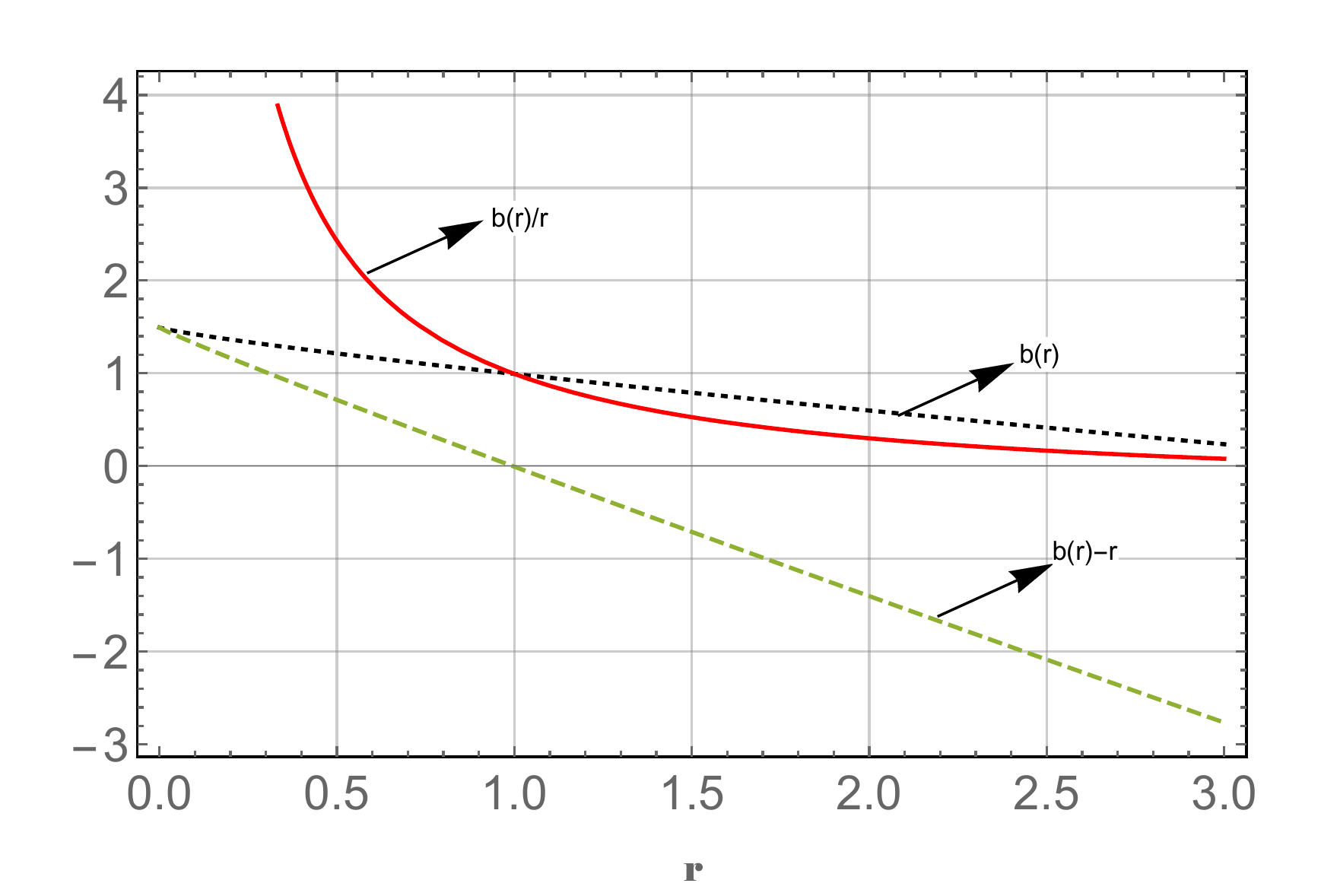}
  \caption{$b(r)$, $b(r)/r$ and $b(r)-r$ as functions of $r$ for $\beta=0.84$, $a=-0.5$ and $r_0=0.995$.}\label{fig0}
\end{figure}

With the purpose of clarifying the metric conditions mentioned above, in Fig.\ref{fig0} below we plot $b(r)$, $b(r)/r$ and $b(r)-r$, for $\beta=0.84$, $a=-0.5$ and $r_0=0.995$.

Using (\ref{16}) in (\ref{12})-(\ref{14}), $\rho$, $p_r$ and $p_t$ read

\begin{widetext}
\begin{equation}\label{17}
\rho=-\frac{a \beta r_0 \left(\frac{r}{r_0}\right)^{\beta } \left\{-12 \pi  \left[r^3-2 a \alpha  \beta  r_0 \left(\frac{r}{r_0}\right)^{\beta }\right]+5 a \alpha  \beta  \lambda  r_0 \left(\frac{r}{r_0}\right)^{\beta }-2 \lambda  r^3\right\}}{3 (\lambda +4 \pi ) (\lambda +8 \pi ) r^6},
\end{equation}

\begin{multline}\label{18}
p_r=\frac{r_0 \left\{-a^2 \alpha  \beta  r_0 [(7 \beta -12) \lambda +24 \pi  (\beta -2)] \left(\frac{r}{r_0}\right)^{2 \beta }-a \left(\frac{r}{r_0}\right)^{\beta } \left[\lambda  \left(12 (a-1) \alpha  \beta  r_0-(\beta -3) r^3\right)+12 \pi  \left((4a-4) \alpha  \beta r_0+r^3\right)\right]\right\}}{3 (\lambda +4 \pi ) (\lambda +8 \pi ) r^6}\\
+\frac{r_0 \left[3 (a-1) (\lambda +4 \pi ) r^3\right]}{3 (\lambda +4 \pi ) (\lambda +8 \pi ) r^6},
\end{multline}

\begin{multline}\label{19}
p_t=\frac{r_0 \left\{-2 a^2 \alpha  \beta r_0 [(\beta +6) \lambda +24 \pi ] \left(\frac{r}{r_0}\right)^{2 \beta }+a \left(\frac{r}{r_0}\right)^{\beta } \left[\lambda  \left(12 (a-1) \alpha  \beta r_0-(\beta -3) r^3\right)+12 \pi  \left(4 (a-1) \alpha  \beta r_0-(\beta -1) r^3\right)\right]\right\}}{6 (\lambda +4 \pi ) (\lambda +8 \pi ) r^6}
\\-\frac{r_0 \left[3 (a-1) (\lambda +4 \pi ) r^3\right]}{6 (\lambda +4 \pi ) (\lambda +8 \pi ) r^6}.
\end{multline}
\end{widetext}

\begin{figure}[ht!]
\centering
  \includegraphics[width=75mm]{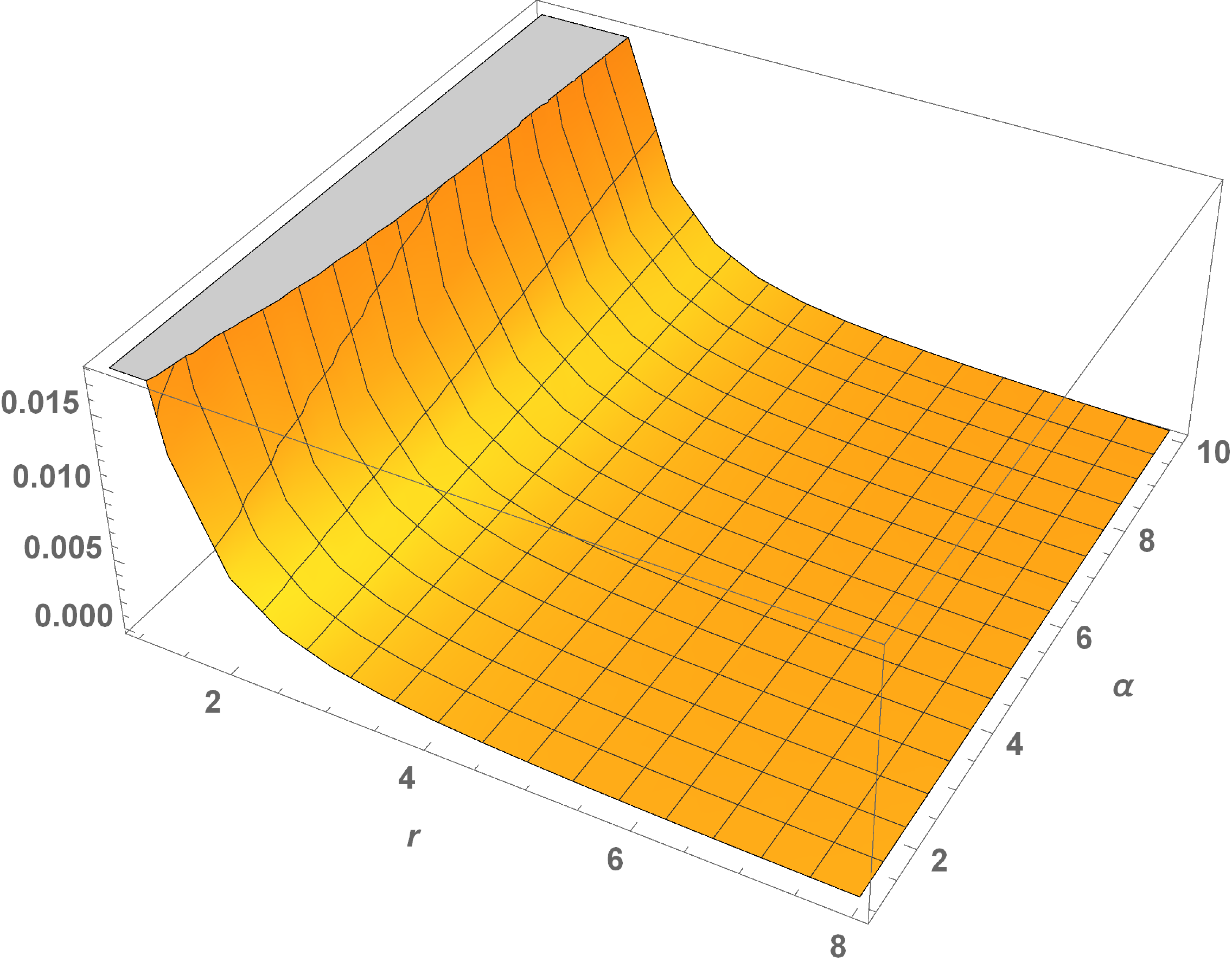}
  \caption{Variation of energy density for $\lambda=-35$ and different $\alpha $.}\label{fig1}
\end{figure}

\begin{figure}[ht!]
\centering
  \includegraphics[width=75mm]{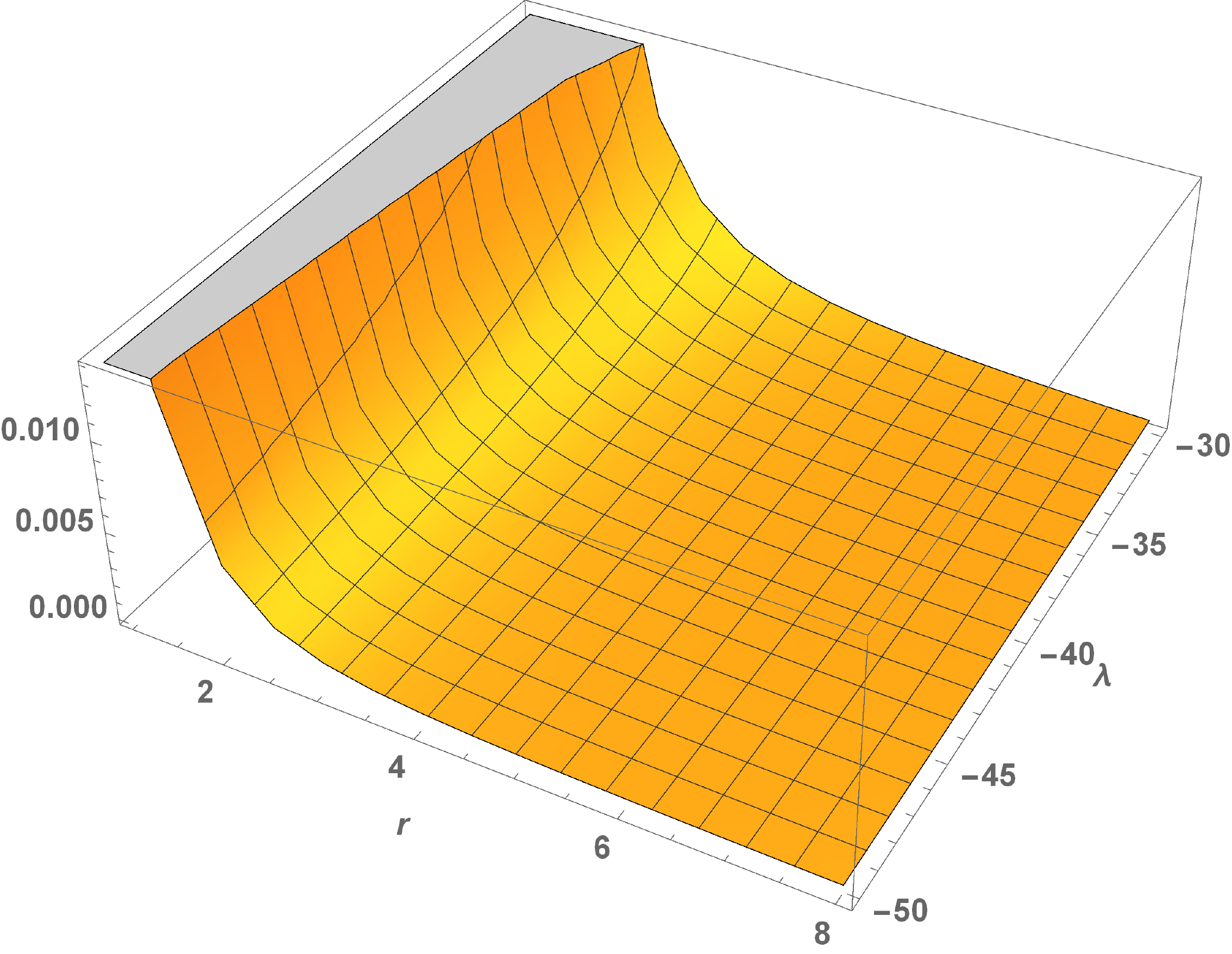}
  \caption{Variation of energy density for $\alpha=5$ and different $\lambda $.}
\label{fig2} 
\end{figure}

\begin{figure}[ht!]
\centering
  \includegraphics[width=75mm]{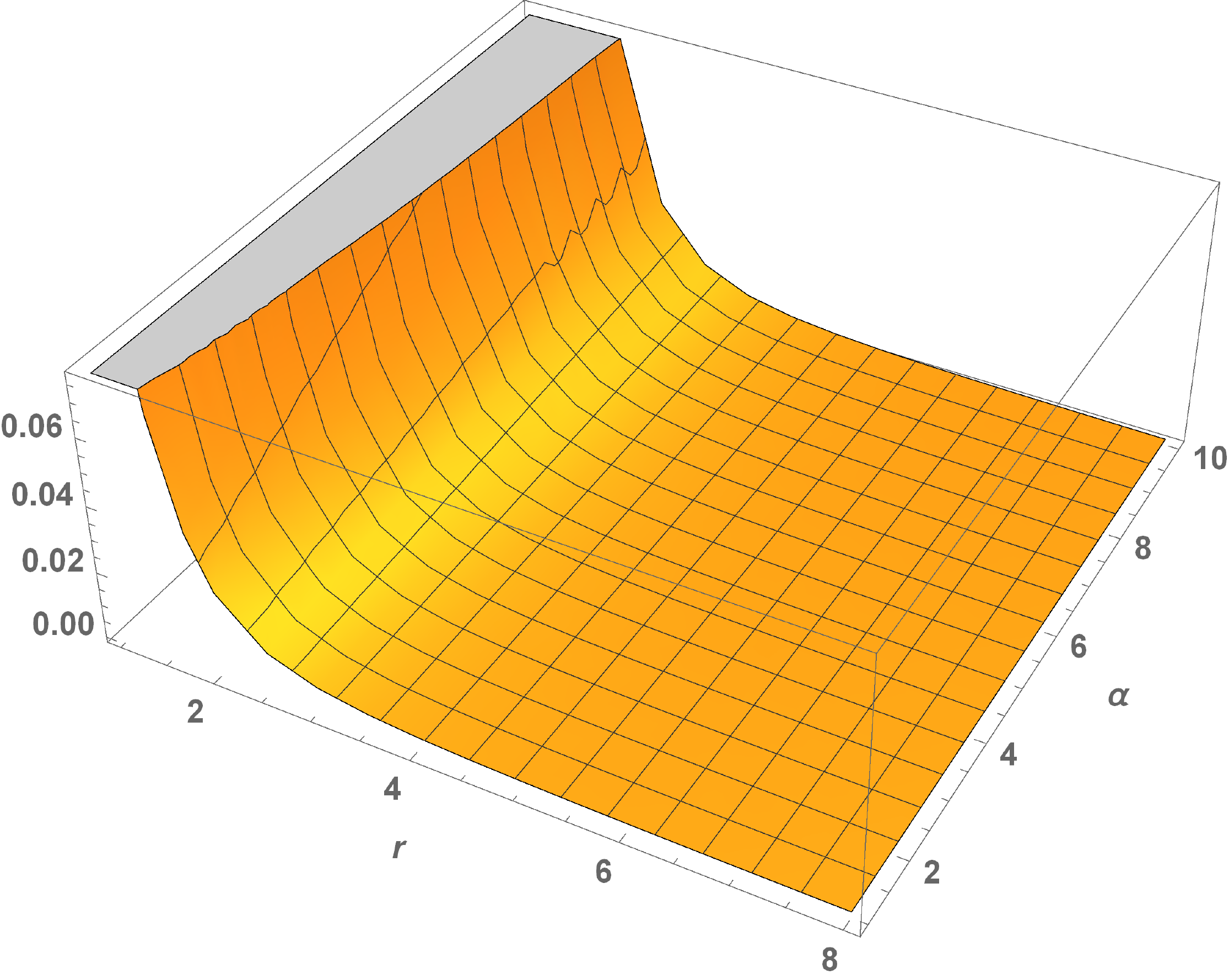}
  \caption{Validity of NEC, $\rho+p_r\geq 0$, for $\lambda=-35$ and different $\alpha $.}\label{fig3}
\end{figure}

\begin{figure}[ht!]
\centering
  \includegraphics[width=75mm]{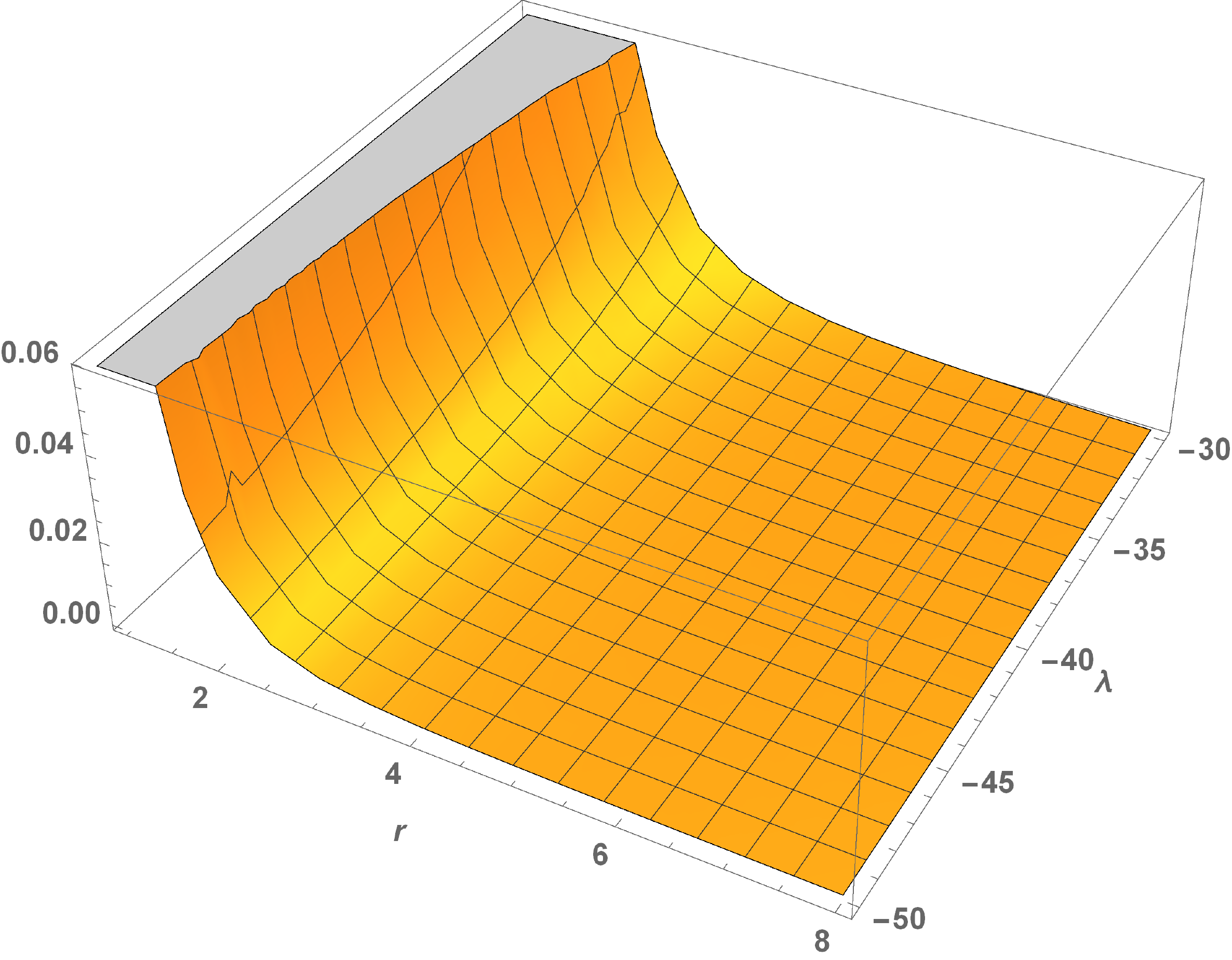}
  \caption{Validity of NEC, $\rho+p_r\geq 0$, for $\alpha=5$ and different $\lambda $.}\label{fig4}
\end{figure}

\begin{figure}[ht!]
\centering
  \includegraphics[width=75mm]{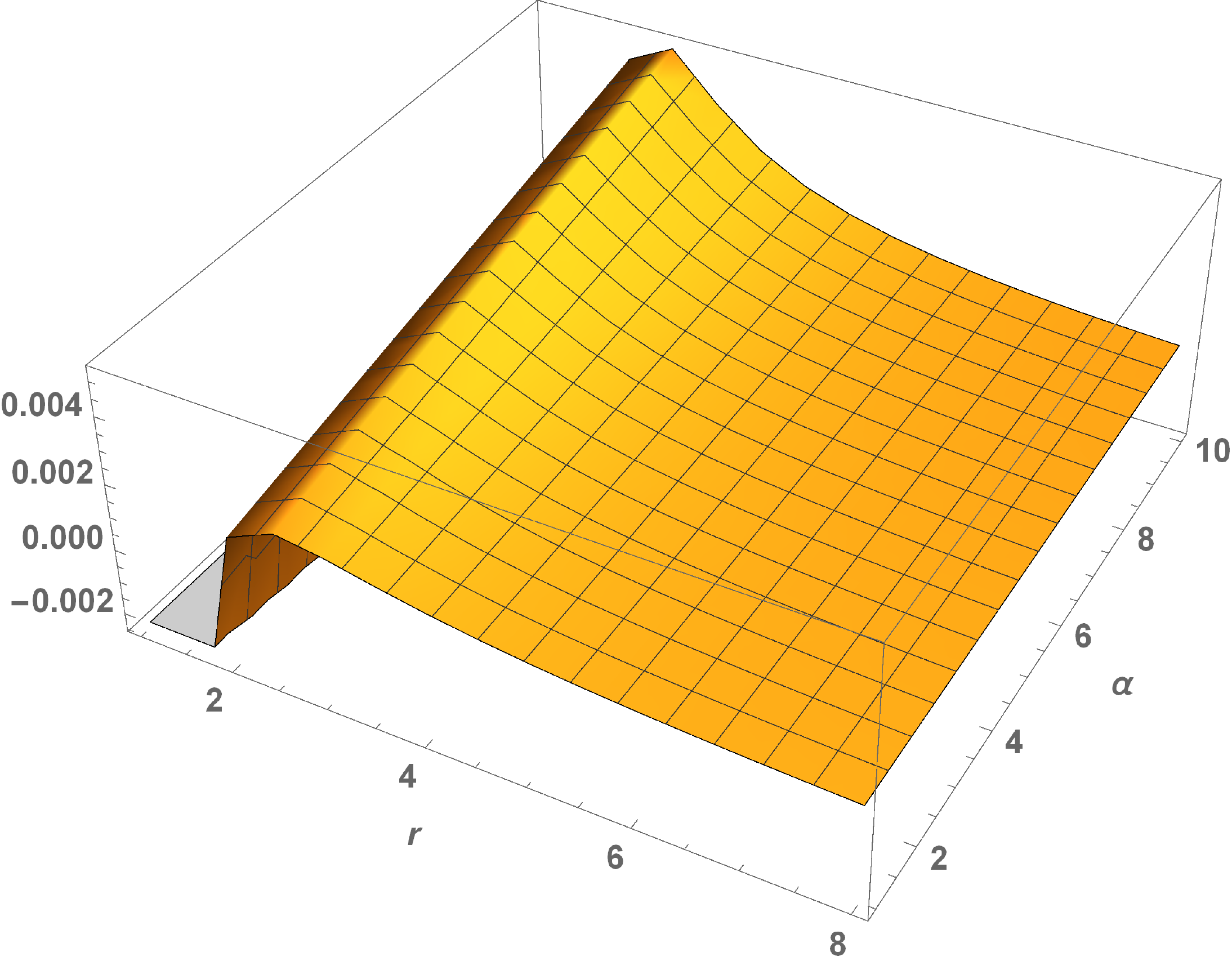}
  \caption{NEC, $\rho+p_t\geq 0$, for $\lambda=-35$ and different $\alpha $.}\label{fig5}
\end{figure}

\begin{figure}[ht!]
\centering
  \includegraphics[width=75mm]{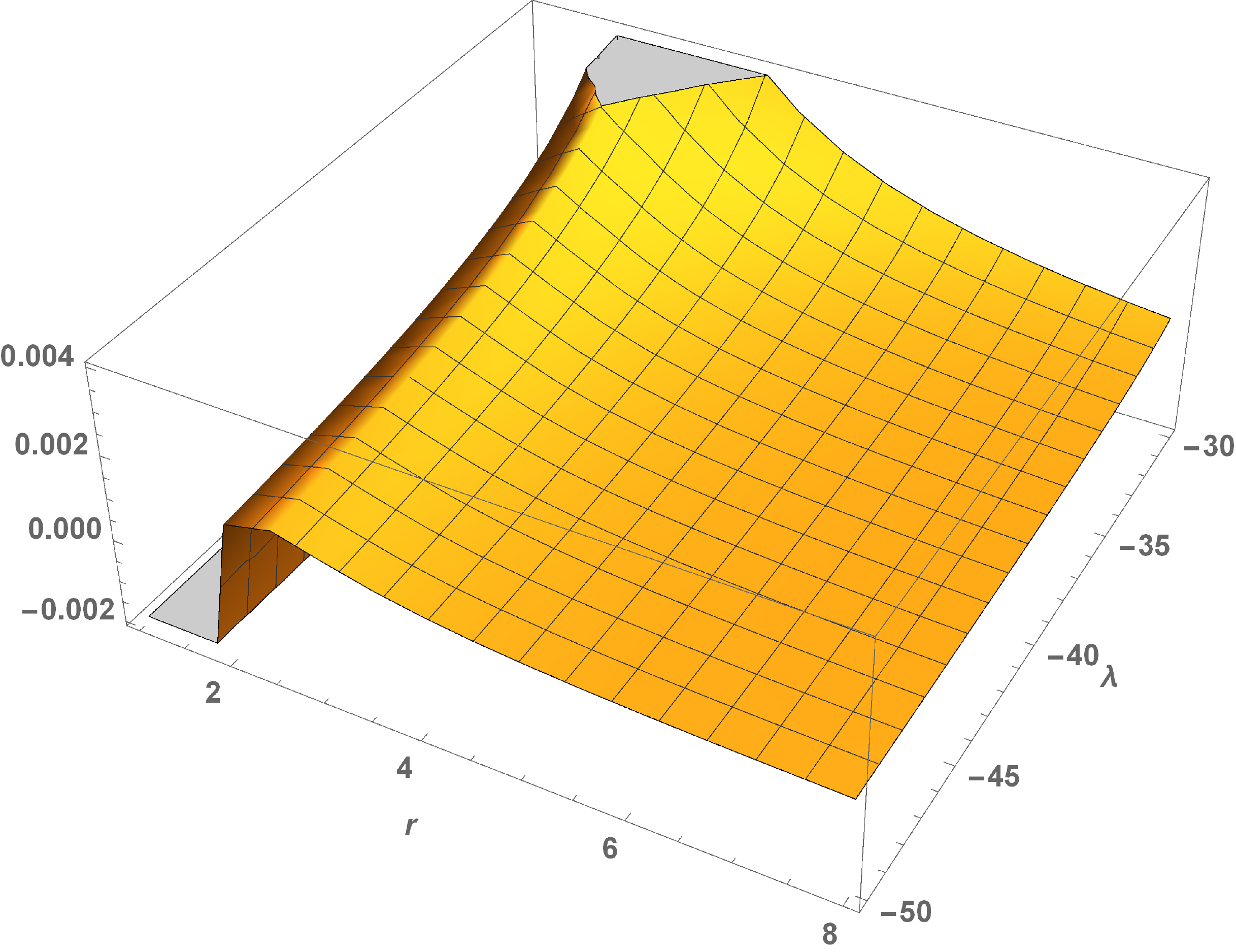}
  \caption{NEC, $\rho+p_t\geq 0$, for $\alpha=5$ and different $\lambda $.}\label{fig6}
\end{figure}

\begin{figure}[ht!]
\centering
  \includegraphics[width=75mm]{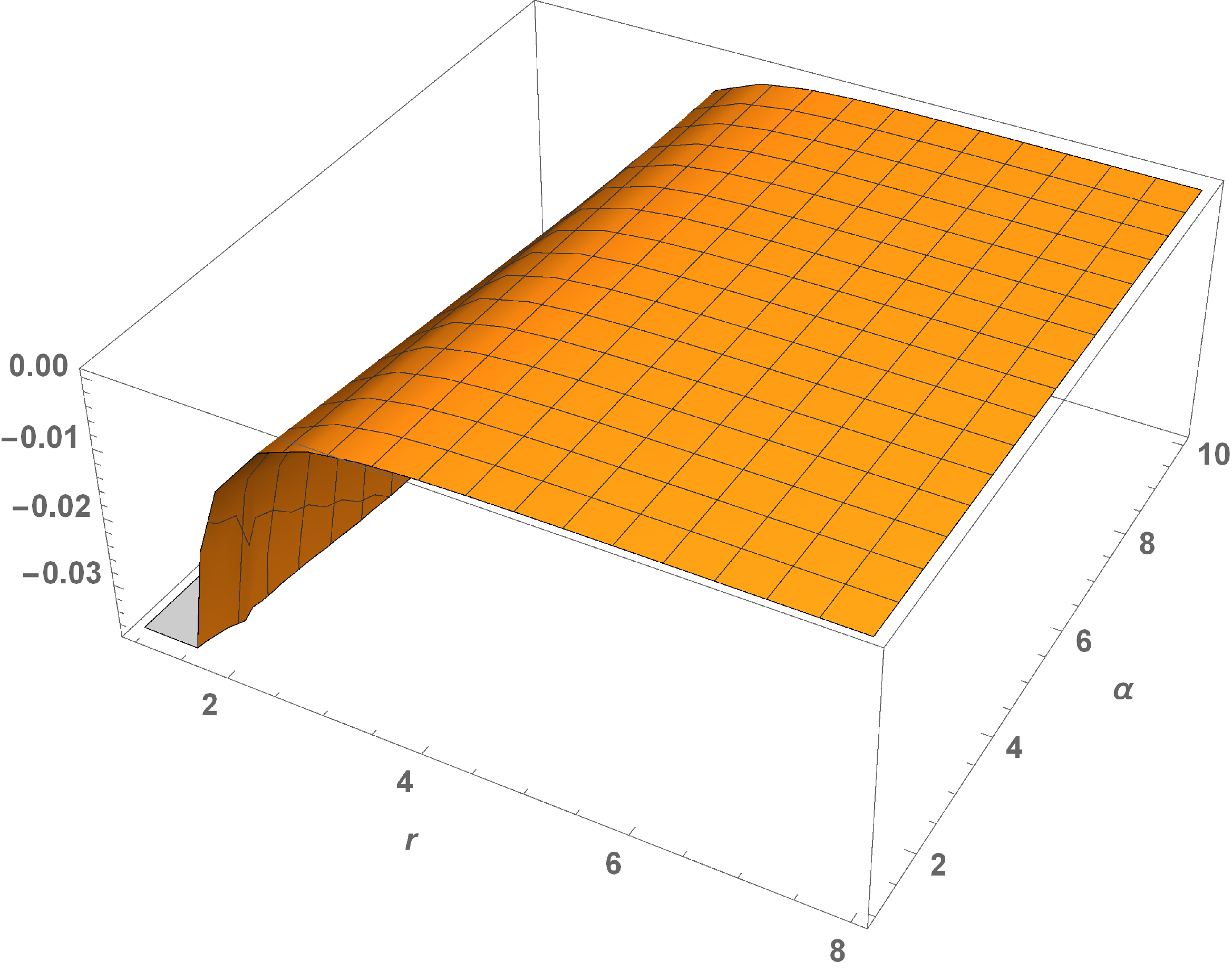}
  \caption{Violation of dominant energy condition (DEC), $\rho\geq \vert p_r\vert$, for $\lambda=-35$ and different $\alpha $.}\label{fig7}
\end{figure}

\begin{figure}[ht!]
\centering
  \includegraphics[width=75mm]{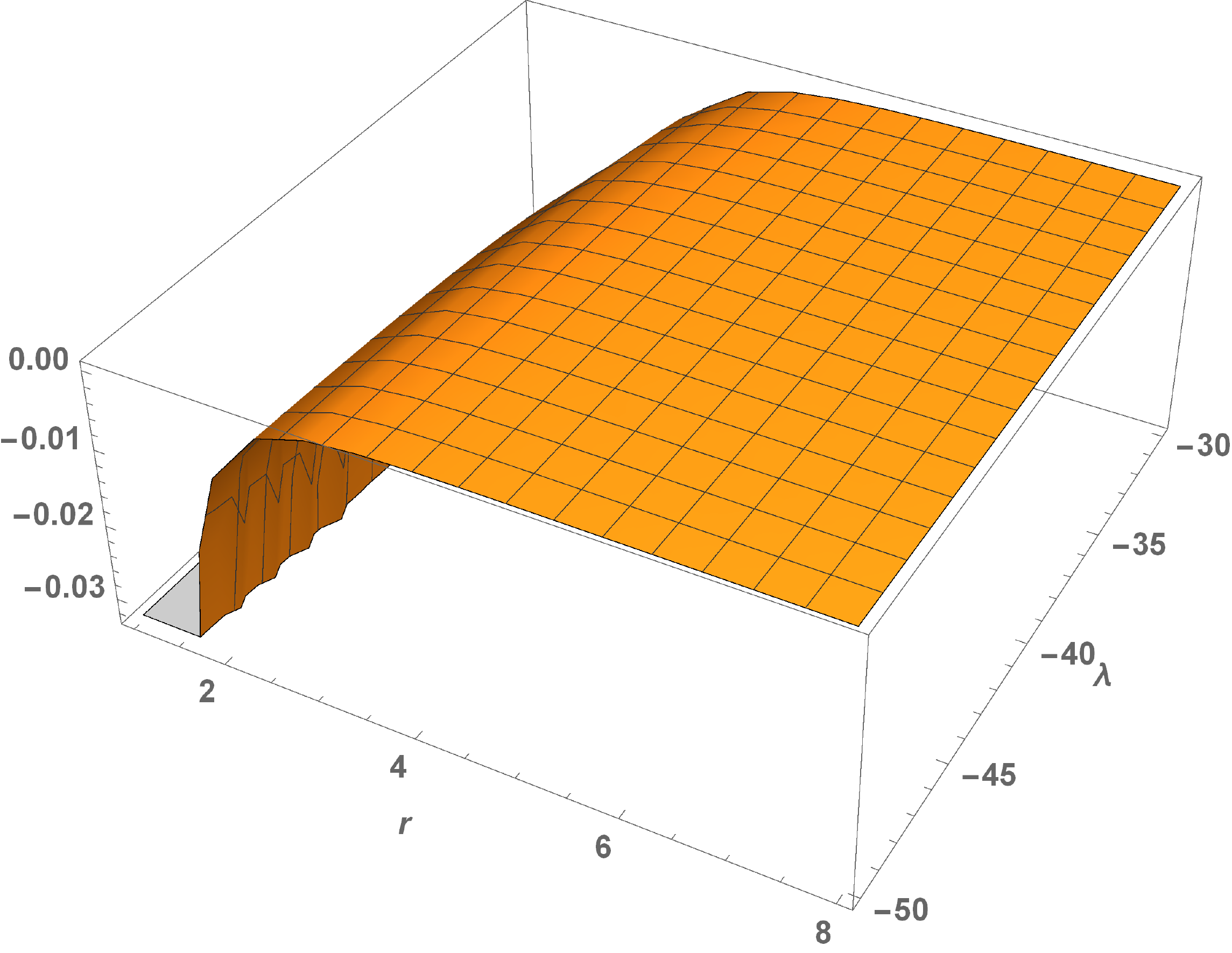}
  \caption{Violation of DEC, $\rho\geq \vert p_r\vert$, for $\alpha=5$ and different $\lambda $.}\label{fig8}
\end{figure}

\begin{figure}[ht!]
\centering
  \includegraphics[width=75mm]{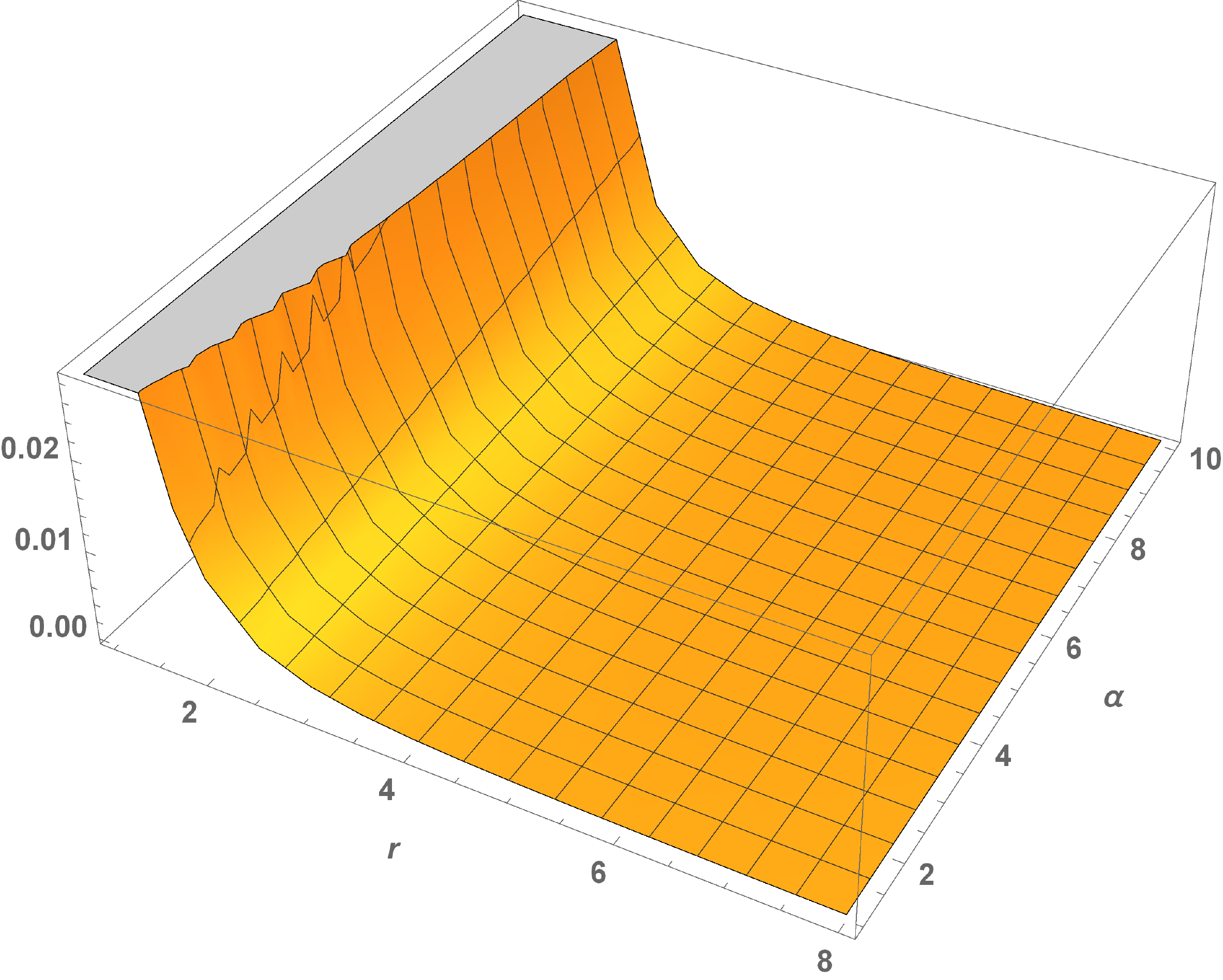}
  \caption{Validation of DEC, $\rho\geq \vert p_t\vert$, for $\lambda=-35$ and different $\alpha $.}\label{fig9}
\end{figure}

\begin{figure}[ht!]
\centering
  \includegraphics[width=75mm]{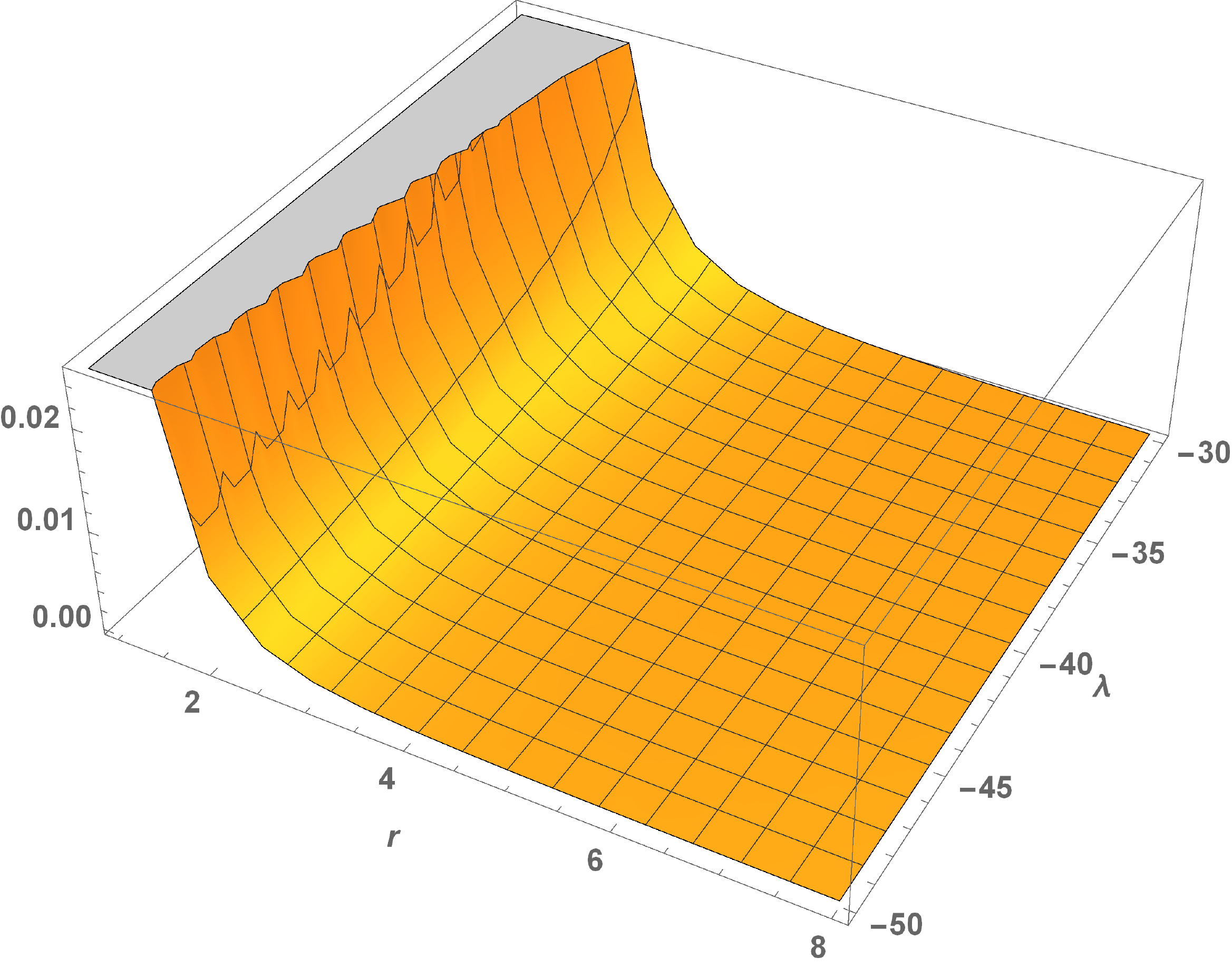}
  \caption{Validation of DEC, $\rho\geq \vert p_t\vert$, for $\alpha=5$ and different $\lambda $.}\label{fig10}
\end{figure}

\begin{figure}[ht!]
\centering
  \includegraphics[width=75mm]{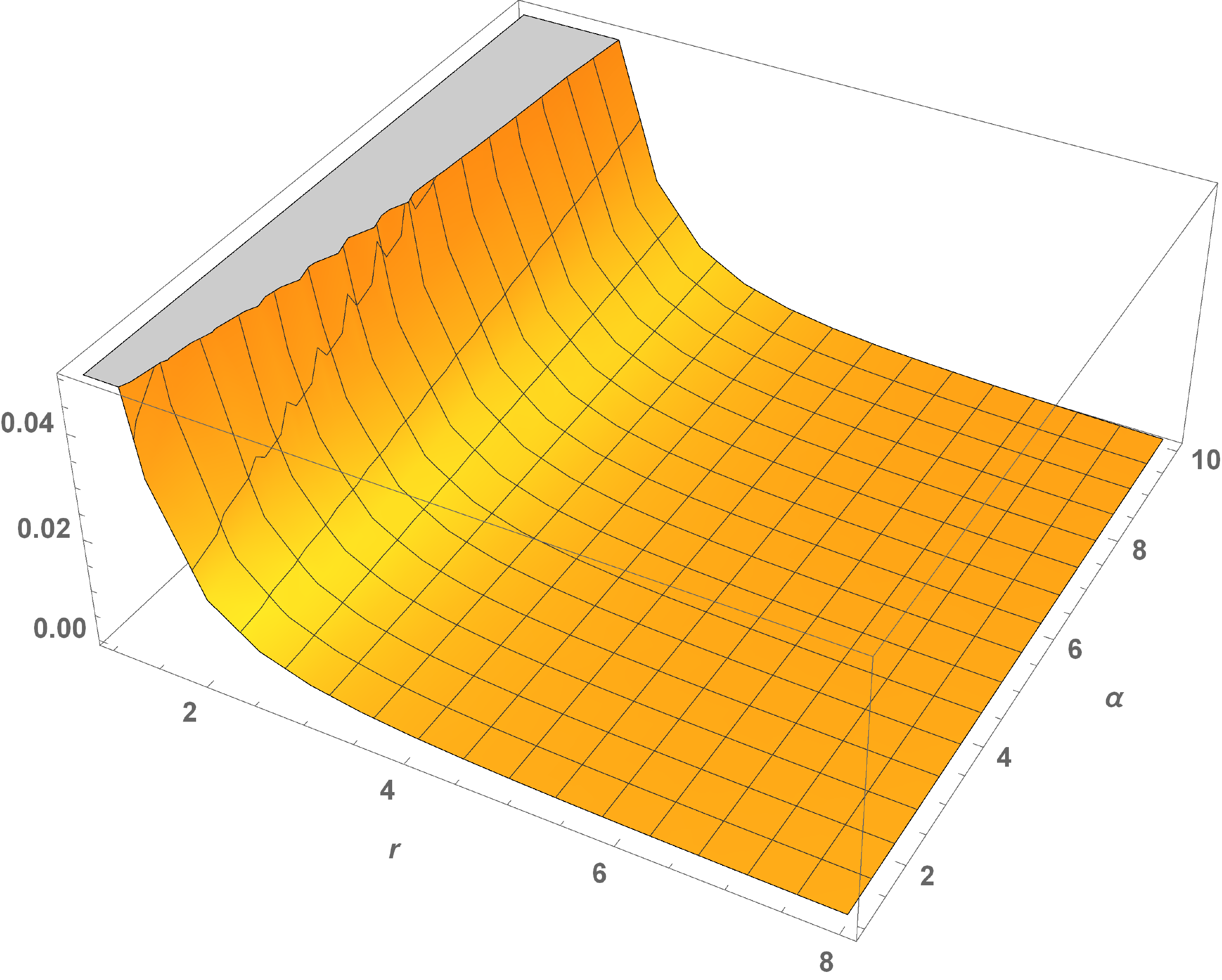}
  \caption{Validation of strong energy condition (SEC), $\rho+p_r+2p_t\geq0$, for $\lambda=-35$ and different $\alpha $.}\label{fig11}
\end{figure}

\begin{figure}[ht!]
\centering
  \includegraphics[width=75mm]{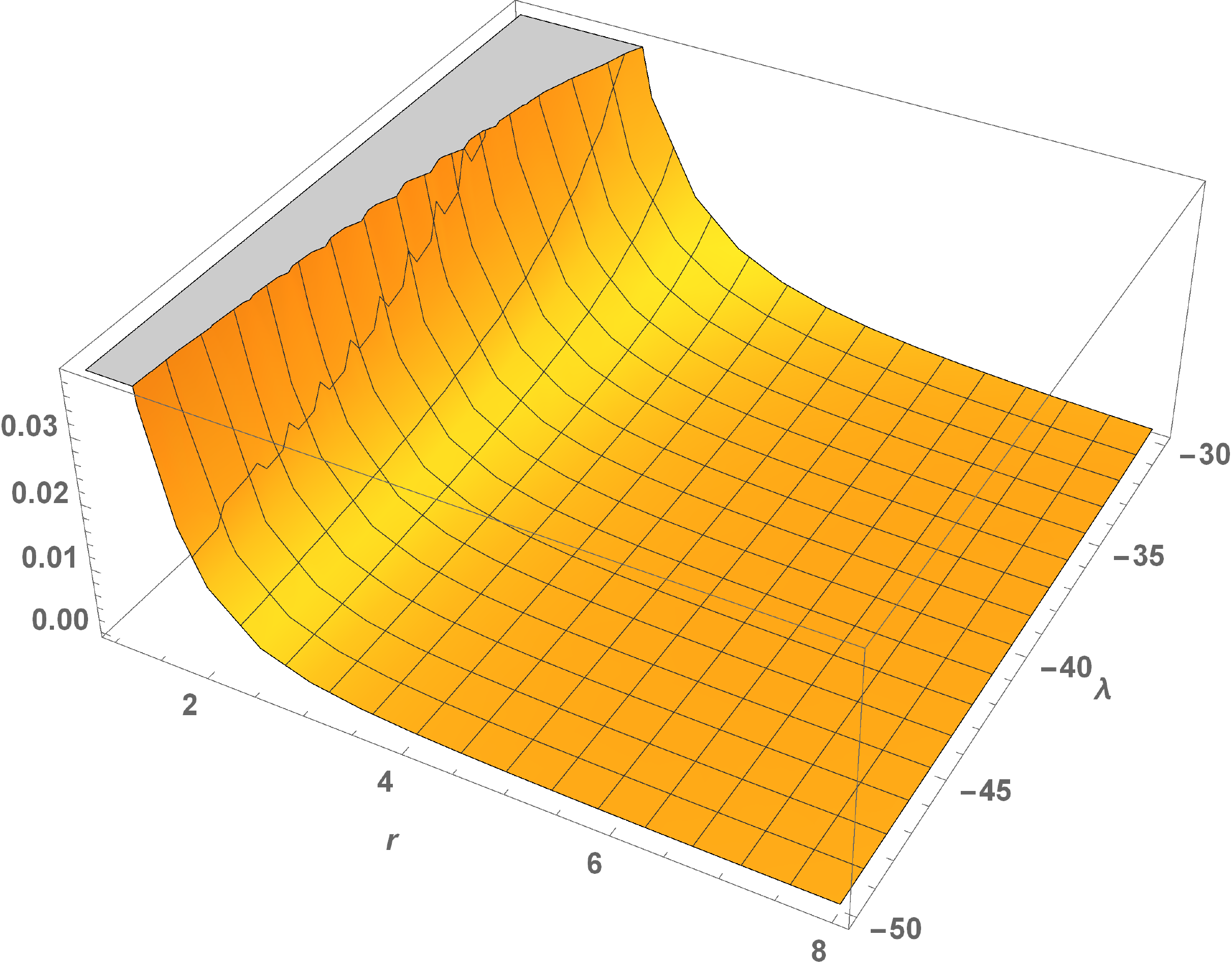}
  \caption{Validation of SEC, $\rho+p_r+2p_t\geq0$, for $\alpha=5$ and different $\lambda $.}\label{fig12}
\end{figure}

From the quantities above, we plot the energy density as well as the energy conditions in Figs.\ref{fig1}-\ref{fig12}. In all figures, we take $\beta=0.84$, $a=-0.5$ and $r_0=0.995$.

\section{Discussion}

We have obtained here, for the first time in the literature, WH solutions in the $f(R,T)=R+\alpha R^{2}+\lambda T$ gravity. Such a gravity theory can be seen as the simplest and more natural theory which presents corrections in both the geometrical and material sectors. Although it has already been applied to the study of compact astrophysical objects, yielding valuable results \cite{noureen/2015}-\cite{zubair/2015}, this is the first time WHs are analysed from such a functional form for $f(R,T)$.

The $R^2$-correction only is well motivated by its applications in cosmology and astrophysics \cite{starobinsky/1980}, \cite{ellis/2013}-\cite{koshelev/2016}. The $T$-correction only is inserted with the purpose of considering quantum effects in a gravity theory, or even the existence of imperfect fluids in the universe \cite{harko/2011}. In this way, since WHs material content is described by an anisotropic imperfect fluid, the WH analysis in theories with material corrections is well motivated.

We remark that the $T-$dependence can also account for a non-minimal coupling between geometry and matter, as explored in \cite{ms/2017}. The existence of WH solutions in curvature-matter coupled gravity have already been discussed in the literature, as it can be checked in \cite{Bertolami/2008,Garcia/2010}. The achievement of WH solutions in a non-minimal geometry-matter coupling model in $f(R,T)$ gravity, following the steps in \cite{ms/2017}, shall be reported soon in the literature.

In the present article, our WH analysis was made by considering Eq.(\ref{16}) as the form for the shape function. Such an assumption was also taken in other references, such as \cite{Heydarzade/2015,lobo/2013}. From (\ref{16}), we were able to obtain the material solutions of the WH, i.e., $\rho$, $p_r$ and $p_t$, as Eqs.(\ref{17})-(\ref{19}), respectively. 

On the regard of the matter content of the present WH solutions, we would like to highlight that, departing from what can be seen in the $f(R,T)$ WHs literature \cite{Azizi/2013}-\cite{Yousaf/2017a}, no equation of state (EoS) was assumed in this article. Since the WH EoS is still poorly known, this is a considerable advantage of the present solutions. It is worth to remark that in future, with eventual detections of WHs (check \cite{tsukamoto/2012}-\cite{ohgami/2015} for some proposals), one may be able to constraint their EoS.

In possession of Eqs.(\ref{17})-(\ref{19}) we plotted the WH energy conditions, namely NEC, DEC and SEC. We mention that the weak energy condition $\rho\geq0$ obedience can also be appreciated in Figs.\ref{fig1}-\ref{fig2}.

In Figs.\ref{fig3}-\ref{fig4}, the validity of NEC, $\rho+p_r\geq0$, for different $\alpha$ and different $\lambda$, respectively, is plotted. 

Figs.\ref{fig5}-\ref{fig6} also present NEC, but in terms of $p_t$. We see the violation of NEC only for small values of $r$.

For DEC, while Figs.\ref{fig7}-\ref{fig8}, in terms of $p_r$, show its violation for most values of $r$, Figs.\ref{fig9}-\ref{fig10}, in terms of $p_t$, indicate its respectability for all $r$.

Figs.\ref{fig11}-\ref{fig12} show that SEC is respected for the present WH matter content, for different values of $\alpha$ and $\lambda$, respectively.

The respectability of the energy conditions in a wide region or even in the whole WH, which yields non-exotic matter to fill in this object, was attained as a consequence of the extra (or correction) terms of the theory, namely $\alpha R^{2}$ and $\lambda T$. As remarked above, non-exotic matter WHs are non-trivial to be attained and we believe that once the material correction terms, predicted by the $f(R,T)$ theory, are related with the possible existence of imperfect fluids in the universe, this theory may provide a breakthrough in the understanding of WHs. 

Particularly, the physical reasons for the obedience of the energy conditions as a consequence of the extra terms of the theory worth a deeper discussion. Apparently, the cosmological and astrophysical observational issues we face nowadays may be overcome by either an alternative gravity theory or a non-standard EoS to describe the matter content concerned. Let us take the dark energy problem, for example. It is well known that the present universe undergoes an accelerated phase of expansion \cite{Perlmutter/1999,Riess/1998}. The counter-intuitive effect of acceleration may be described either by an exotic EoS for the matter filling the universe, namely $\mathcal{P}\sim-\rho$ \cite{hinshaw/2013}, or by alternative gravity models, as it can be checked in \cite{Starobinsky/2007}, for example. The recently detected massive pulsars \cite{demorest/2010,antoniadis/2013} can also be attained by particular EoS \cite{ozel/2016,fortin/2015} or alternative gravity \cite{mam/2016}.

The same picture can be visualized for the WHs case, i.e., the extra degrees of freedom of an extended theory of gravity may also allow WHs to be filled by non-exotic EoS matter, departing from the General Relativity case. The $T$-dependence of the $f(R,T)$ theory may characterize the first steps in describing quantum effects in a gravity theory \cite{harko/2011,ms/2017} and such a description, which is missing in General Relativity, can explain the energy conditions obedience. 

As a forthcoming work, we can check the viability of evolving WHs within the present formalism, as well as possible observational signatures these objects may imprint in the galactic halo, as they evolve \cite{rahaman/2014}-\cite{li/2014}. With the purpose of further checking the reliability of the present theory, a cosmological model from the $f(R,T)=R+\alpha R^{2}+\lambda T$ formalism shall also be reported soon in the literature.

\begin{acknowledgements}
PKS and PS acknowledges DST, New Delhi, India for providing facilities through DST-FIST lab, Department of Mathematics, where a part of this work was done. PHRSM would like to thank S\~ao Paulo Research Foundation (FAPESP), grant 2015/08476-0, for financial support. The authors also thank the referee for the valuable suggestions, which improved the presentation of the present results.
\end{acknowledgements}



\end{document}